\documentclass[10pt,journal,twocolumn,pagenumbers]{IEEEtran}

\makeatletter
\let\NAT@parse\undefined
\makeatother

\usepackage[numbers]{natbib}

\usepackage[cmex10]{amsmath}
\usepackage{amssymb}
\usepackage{xy}
\usepackage[vflt]{floatflt}
\usepackage[dvips]{graphicx}
\DeclareGraphicsExtensions{.eps}

\usepackage{hyperref}

\newfont{\hiera}{cmsy10 scaled 2488} 
\newfont{\hierb}{cmsy10 scaled 1728}
\newfont{\hierc}{cmsy10 scaled 1200}

\newcommand{\Bigast}{
\mathop{\vphantom{\sum}\lower2.5pt\hbox{\hiera\char3}}}%

\newcommand{\Bigtimes}{
\mathop{\vphantom{\sum}\lower2.5pt\hbox{\hiera\char2}}}%

\def\BibTeX{{\rm B\kern-.05em{\sc i\kern-.025em b}\kern-.08em
    T\kern-.1667em\lower.7ex\hbox{E}\kern-.125emX}}

\newtheorem{theorem}{Theorem}

\newtheorem{definition}[theorem]{Definition}

\begin{document}

\title{\Large Supernodal Analysis Revisited\\
{\large (2009 Re-Release)}
}

\author{{\large Eberhard H.-A.\ Gerbracht
\thanks{This article first appeared in: SMACD'04 Proceedings of the International Workshop on Symbolic Methods and Applications in Circuit Design, Wroclaw, Poland, 23-24 September 2004. Sidney 2004, pp.~113--116. Due to the low distribution of these proceedings, the author has decided to make the article available to a larger audience through the arXiv.}
\thanks{The original research which led to this paper was done while the author was with the Institut f\"ur Netz\-werk\-theorie und Schaltungstechnik, Technische Universit\"at Braunschweig, D-38106 Braunschweig, Germany. The preparation of this paper and its presentation at the SMACD'04 workshop was not supported by any public funding.}
\thanks{The author's current (as of March~9th, 2009) address is Bismarckstra\ss e 20, D-38518 Gifhorn, Germany. Current e-mail: \tt{e.gerbracht@web.de}}}}

\maketitle
\pagestyle{empty}

\bigskip

\begin{abstract}
In this paper we show how to extend the known algorithm of nodal analysis in such a way that, in the case of circuits without nullors and controlled sources (but allowing for both, independent current and voltage sources), the system of nodal equations describing the circuit is partitioned into one part, where the nodal variables are explicitly given as linear combinations of the voltage sources and the voltages of certain reference nodes, and another, which contains the node variables of these reference nodes only and which moreover can be read off directly from the given circuit. Neither do we need preparational graph transformations, nor do we need to introduce additional current variables (as in MNA). Thus this algorithm is more accessible to students, and consequently more suitable for classroom presentations.
\end{abstract}

{\small
\smallskip

\smallskip

}

\section{Introduction}
\noindent 
It is a well known fact -- already taught in most undergraduate courses on circuit theory \cite{Balabanian} -- that,
when the node equations for a standard {\sl Nodal Analysis (NA)} of a linear circuit containing only admittances and independent current sources are set up in matrix form
\begin{equation*}
{\bf Y}_n {\bf v}_n = {\bf J}_n,
\end{equation*}
where ${\bf v}_n$ denotes the vector of node voltages $v_{\xy*+{\scriptstyle{1}}*\cir{}\endxy},\dots,v_{\xy*+{\scriptstyle{n}}*\cir{}\endxy}$ of those nodes different from some fixed {\em reference node} ${\xy*+{0}*\cir{}\endxy},$ 
the entries of the {\em node-admittance matrix} ${\bf Y}_n$ and the {\em node current-source vector} ${\bf J}_n$ can be directly read off from the circuit itself. I.e., each diagonal term of  ${\bf Y}_n$ in position $(i,i)$ is given by the sum of admittances incident with the node ${\xy*+{i}*\cir{}\endxy}$, each off-diagonal term in position $(i,j),$ $i\ne j,$ is described by the negative of the sum of admittances connecting the nodes ${\xy*+{i}*\cir{}\endxy}$ and ${\xy*+{j}*\cir{}\endxy};$ the $i$-th entry of the vector ${\bf J}_n$ is the sum of all independent currents leaving or entering the node ${\xy*+{i}*\cir{}\endxy}$ with a plus sign attached only to those currents directed toward the node and a minus sign to all the others.

As is equally well known, while it is easy to extend nodal analysis to deal with circuits, which furthermore contain voltage controlled current sources or nullors, massive problems arise, when voltage sources of any kind have to be taken into consideration, as well. Although this obstacle has been basically overcome by the invention of the {\sl Modified Nodal Analysis (MNA)}, which gives a universal method for any kind of linear circuit, with good reasons most teachers of Electrical Engineering seem to be very reluctant to confront their students with the MNA-algorithm, especially in undergraduate courses.   

Accordingly, several authors (\cite{SommerSuper,ChenDavis}) have proposed another alternative, the so-called {\sl Supernodal Ana\-ly\-sis (SNA)}, which seems to be more accessible to students and thus has been incorporated into existing undergraduate and graduate courses\footnote{Since the first publication of this article, the lecture notes \cite{SommerVL} seem not to be publicly available anymore. Nevertheless, those contents of the course which were relevant for this paper have been incorporated into \cite{SommerGST}.} (\cite{Davis,SommerVL,GerbrachtVL}):
Starting with a linear circuit with admittances and all kinds of independent sources, the initial set of equations can be reduced to a smaller set; these resulting SNA-equations again can be described in matrix form as
\begin{equation*}
\widehat{\bf Y}_N \widehat{\bf v}_N = \widehat{\bf J}_N,
\end{equation*}
where $\widehat{\bf v}_N$ is a vector of selected node voltages, $\widehat{\bf Y}_N$ is a matrix of admittances and $\widehat{\bf J}_N$ consists of suitably chosen linear combinations of the independent sources. Although in the literature (\cite{ChenDavis,Davis}) instructions are given how to calculate the entries  of  $\widehat{\bf Y}_N$ and $\widehat{\bf J}_N,$  as far as we know, no algorithm has been developed, yet, which in analogy to standard nodal analysis allows one to directly read them off from the circuit. This paper was written to remedy this situation.

\smallskip
{\em Remark:}\;
Throughout this paper, to keep notation as simple as possible, while we freely talk about circuits with admittances, all the examples will be linear circuits con\-sider\-ed in the time domain, which besides independent sources consist of resistors with positive conductances (symbolized by capital letters), controlled sources and/or nullors. The experienced reader will know how to generalize the results, which will be presented, to other linear circuits containing inductors and capacitors.

The only notational convention we will strictly adhere to is using boldface letters for vectors and matrices. 

\smallskip
Without loss of generality (cp.\ \cite{ChuaDesoerKuh}, 1.5.3) we demand that all circuits under consideration are connected.


\section{Supernodal Analysis of Linear Circuits with Independent Sources}

To keep matters simple at the beginning, in this section we will only consider circuits without controlled sources and nullors.

The basis of our discussion is the concept of a {\em super\-node.} The definition, we will give, slightly differs both from the one presented in \cite{SommerSuper}, as well as from the one in \cite{Davis}, chapter 4.2. This was done to streamline the formulation of the ``traditional'' algorithm and to prepare for our improvements.

\smallskip
\begin{definition}
\label{super1}
A subcircuit of a given circuit which is connected, consists only of nodes and (independent) voltage sources, and which is maximal with these two properties\footnote
{I.e.\ it cannot be enlarged without losing one or the other attribute.} is called a {\em supernode.}  
\end{definition}

\smallskip
Let us remark, that by this definition an ordinary node which is not incident with any voltage source is regarded as a supernode, as well. Furthermore, any super\-node defines a cut of the circuit. The {\em contraction\footnote
{in the sense of \cite{Bollobas}, exercise II.4.15.}
} $\widehat{\Gamma}$ of a circuit $\Gamma,$ obtained by contracting all the branches of all supernodes and removal of all resulting loops, is a circuit, which by the initial prerequisite of this section only contains resistors and independent current sources. During the course of this paper we will call $\widehat{\Gamma}$ the {\em contraction along the supernodes.}

If, moreover, one removes all branches associated to current sources from $\widehat{\Gamma},$ the result is the {\em deactivated circuit} in the sense of \cite{ChenDavis,Davis}.

\subsection{The Algorithm -- Proceeding as in Textbooks}

We are now able to adapt the general algorithm of {\em Super\-nodal Analysis} as presented in \cite{SommerSuper,ChenDavis,Davis}, and formulate it in our terminology.

\medskip
\noindent
{\bf Input:} a connected circuit $\Gamma$ with $n+1$ nodes, named ${\xy*+{0}*\cir{}\endxy},$ ${\xy*+{1}*\cir{}\endxy},\dots,$${\xy*+{n}*\cir{}\endxy},$ consisting only of independent sources and resistors. 
($*$ Node ${\xy*+{0}*\cir{}\endxy}$ will be our {\em global reference node (ground/datum)}, and we set $v_{\xy*+{\scriptstyle{0}}*\cir{}\endxy} = 0.$   $*$)

\smallskip
\noindent
{\bf Output:} 
\begin{enumerate}
\item a set of equations for the node voltages $v_{\xy*+{\scriptstyle{0}}*\cir{}\endxy},$ $v_{\xy*+{\scriptstyle{1}}*\cir{}\endxy},\dots,v_{\xy*+{\scriptstyle{n}}*\cir{}\endxy},$ which completely describes the circuit $\Gamma.$
\item a subset of node voltages together with a {\em reduced system} of equations (i.e.\ equations containing these variables only), the solution of which directly leads to the solution of the whole system. 
\end{enumerate}
 
\medskip
\noindent
\begin{enumerate}
  \item Initialization
  \begin{enumerate}
     \item
     Assign node voltages.
     \item
     Identify all supernodes and mark them;
     let $N+1$ be the number of nodes in the contraction along the super\-nodes $\widehat{\Gamma}.$
     \item
     Within each supernode define one node as the {\em local reference node} of the supernode (a supernode consisting of a single node only is its own reference node). The global reference node should be chosen to be a local reference node\footnote
{Clearly the global reference node need not be chosen before this step of the algorithm.}.
  \end{enumerate}
  \item
  \label{algoVEq}
  For each supernode, express the node voltages within in terms of the node voltage of its local reference node and the values of the voltage sources, it encompasses.
  \item
  \label{algoKirchhoff}
  For each supernode, write one KCL equation in terms of the voltages of the local reference nodes; leave out, as usual, the one for the supernode containing the global reference node.
\end{enumerate}

\medskip
For those, who are not accustomed, yet, with supernodal analysis, we exemplify the workings of the above algorithm:

\begin{center}
\vskip -0.2cm
{\includegraphics[width=\linewidth,clip,keepaspectratio]{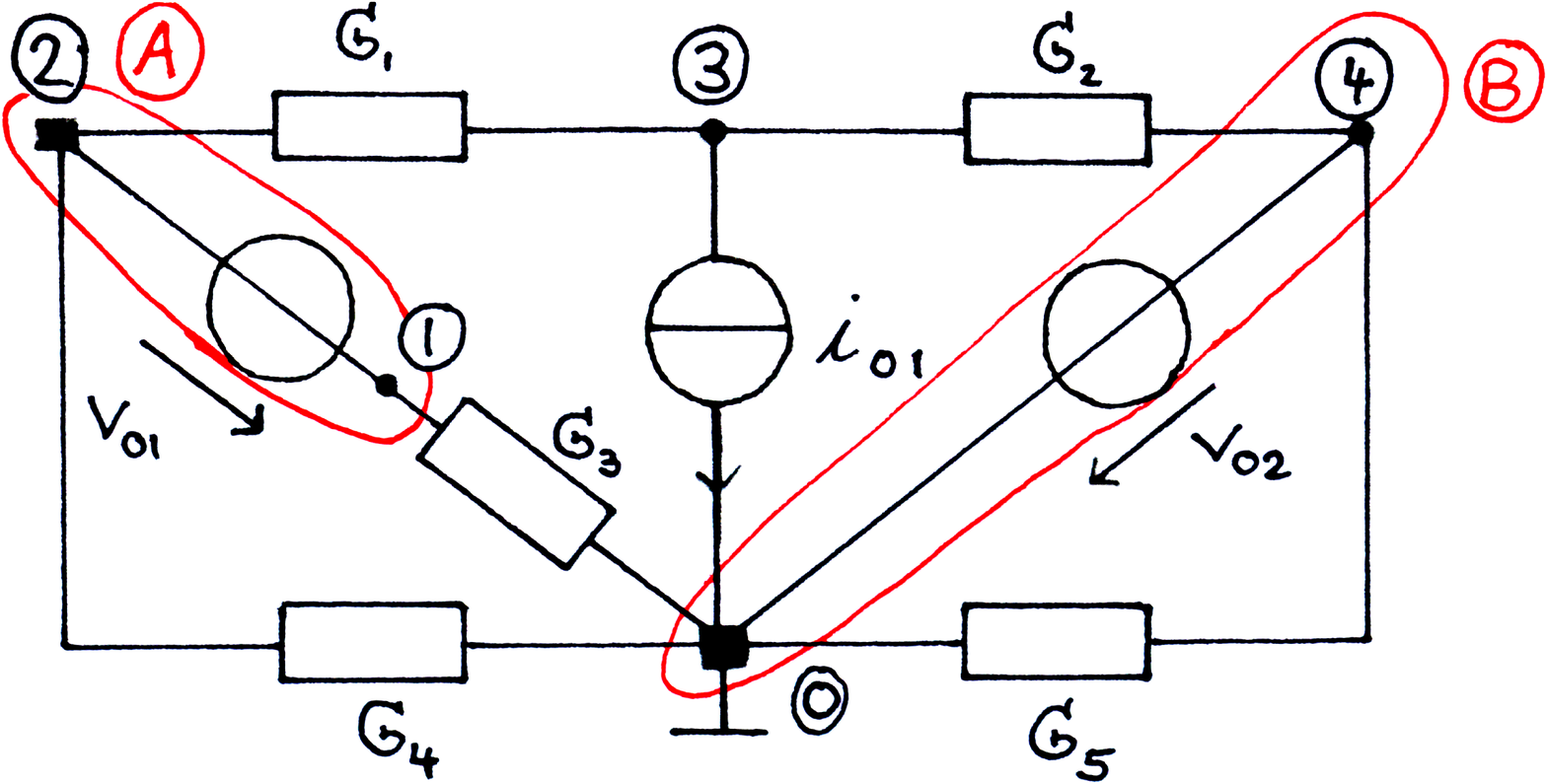}}
\vskip -3truecm
\end{center}
The two voltage sources of this circuit give raise to two supernodes ${\xy*+{\scriptstyle{A}}*\cir{}\endxy}$ and ${\xy*+{\scriptstyle{B}}*\cir{}\endxy}$ consisting of the nodes ${\xy*+{1}*\cir{}\endxy},$${\xy*+{2}*\cir{}\endxy}$ and ${\xy*+{0}*\cir{}\endxy},$${\xy*+{4}*\cir{}\endxy},$ respectively. While we can freely choose ${\xy*+{2}*\cir{}\endxy}$ as the local reference node for the supernode ${\xy*+{\scriptstyle{A}}*\cir{}\endxy},$ the local reference node of ${\xy*+{\scriptstyle{B}}*\cir{}\endxy}$ is supposed to be the node ${\xy*+{0}*\cir{}\endxy}.$

With node voltages being introduced, the internal struct\-ure of the supernodes implies
\begin{equation}
\label{supernode}
\begin{pmatrix}
v_{\xy*+{\scriptstyle{0}}*\cir{}\endxy}\\
v_{\xy*+{\scriptstyle{1}}*\cir{}\endxy}\\
v_{\xy*+{\scriptstyle{4}}*\cir{}\endxy}
\end{pmatrix}
=
\begin{pmatrix}
0\\
v_{\xy*+{\scriptstyle{2}}*\cir{}\endxy} - v_{01}\\
v_{\xy*+{\scriptstyle{0}}*\cir{}\endxy} + v_{02}\\
\end{pmatrix}
=
\begin{pmatrix}
0\\
v_{\xy*+{\scriptstyle{2}}*\cir{}\endxy} - v_{01}\\
v_{02}
\end{pmatrix},
\end{equation}
where we have partially solved this system of equations for the voltages of those nodes, which are not reference nodes.

Kirchhoff equations for supernodes ${\xy*+{\scriptstyle{A}}*\cir{}\endxy}$ and ${\xy*+{3}*\cir{}\endxy}$ give:
\begin{alignat*}{5}
G_1(v_{\xy*+{\scriptstyle{2}}*\cir{}\endxy}-v_{\xy*+{\scriptstyle{3}}*\cir{}\endxy}) &\; +\; & 
G_3(v_{\xy*+{\scriptstyle{1}}*\cir{}\endxy}-v_{\xy*+{\scriptstyle{0}}*\cir{}\endxy}) &\; +\; &
G_4(v_{\xy*+{\scriptstyle{2}}*\cir{}\endxy}-v_{\xy*+{\scriptstyle{0}}*\cir{}\endxy}) 
&\; =\; 0,\\
G_1(v_{\xy*+{\scriptstyle{3}}*\cir{}\endxy}-v_{\xy*+{\scriptstyle{2}}*\cir{}\endxy}) &\; +\; & 
G_2(v_{\xy*+{\scriptstyle{3}}*\cir{}\endxy}-v_{\xy*+{\scriptstyle{4}}*\cir{}\endxy}) &\; +\; &
\quad\quad i_{01}
&\; =\; 0.
\end{alignat*}

Finally, by using (\ref{supernode}) and collecting all currents and voltages resulting from independent sources on the right hand side, we are led to the following system of equations for the voltages of the local reference nodes: 
\begin{equation}
\label{exfinal}
\begin{pmatrix}
G_1+G_3+G_4 & - G_1 \\
- G_1 & G_1+G_2\\
\end{pmatrix}
\cdot
\begin{pmatrix}
v_{\xy*+{\scriptstyle{2}}*\cir{}\endxy}\\
v_{\xy*+{\scriptstyle{3}}*\cir{}\endxy}\\
\end{pmatrix}
=
\begin{pmatrix}
0 &+& G_3\cdot v_{01}\\
-i_{01} &+& G_2\cdot v_{02}\\
\end{pmatrix}
\end{equation}

This reduced system is completely decoupled from (\ref{supernode}). Furthermore its solution together with (\ref{supernode}) describes all the node voltages of our example circuit.


\subsection{On the results of the SNA algorithm in general}

The preceding example gives rise to some observations, which easily generalize to theorems for arbitrary circuits built from admittances and independent sources. Let us suppose that within each supernode one local reference node has already been chosen.

\smallskip
\begin{theorem}
If a supernode carries the structure of a tree then the voltage of any node within the supernode is uniquely expressible as the voltage of the local reference plus the sum, relative to the path orientation, of voltages of those independent sources along the unique path from the reference node to the node under consideration.
\end{theorem}

\smallskip
Thus, step \ref{algoVEq} of the algorithm can be successfully carried out, iff either every supernode is a tree, or any loop of independent voltage sources within any supernode satisfies KVL (which can be easily checked by identifying a spanning tree within each supernode). 

For the sake of completeness, we note that branch voltages along admittances which at both ends are incident with the same supernode can be determined without any calculations.

\smallskip
\begin{theorem}
If one of the above assumptions holds, the system of equations resulting from step \ref{algoKirchhoff} of the algorithm of supernodal analysis can be reduced to
\begin{equation}
\widehat{\bf Y}_N \widehat{\bf v}_N = \widehat{\bf J}_N,
\end{equation}
where $\widehat{\bf v}_N$ is the vector of voltages of local reference nodes, $\widehat{\bf Y}_N$ is the node-admittance matrix of the contraction along the supernodes $\widehat{\Gamma}$ and $\widehat{\bf J}_N$ is a vector of currents induced by the independent sources, which will be specified more precisely below.
\end{theorem}
\subsection{Reading off the Entries from the Circuit}

Since $\widehat{\bf Y}_N$ is the node-admittance matrix of the contraction $\widehat{\Gamma},$ there is a simple rule how to fill in the entries just by inspection.

\smallskip
{\bf Rule 1:}
Each element on the main diagonal at position $(A,A)$
 is the sum of the admittances incident at only one end\footnote{
As noted above, those admittances incident with only one super\-node at both ends have to be neglected.
} 
with the supernode ${\xy*+{\scriptstyle{A}}*\cir{}\endxy}.$ The off-diagonal elements at position $(A,B)$
 is the negative of the sum of those admittances connecting supernodes ${\xy*+{\scriptstyle{A}}*\cir{}\endxy}$ and ${\xy*+{\scriptstyle{B}}*\cir{}\endxy}.$

\smallskip
In the future -- for obvious reasons -- when we talk about admittances ``incident with a supernode'' we will only consider those, that are incident at only one end.

Without loss of generality, let us assume that all super\-nodes of $\Gamma$ are trees. By this additional assumption paths within each supernode are unique. Thus any admittance $G_k$ connecting two supernodes ${\xy*+{\scriptstyle{A}}*\cir{}\endxy}$ and ${\xy*+{\scriptstyle{B}}*\cir{}\endxy}$ defines a unique oriented path $\gamma_{G_k}^A,$ consisting only of independent voltage sources and the admittance $G_k$ itself, 
as sketched in the figure below:

\medskip
\begin{center}
\vskip -0.3cm
{\includegraphics[width=\linewidth,clip,keepaspectratio]{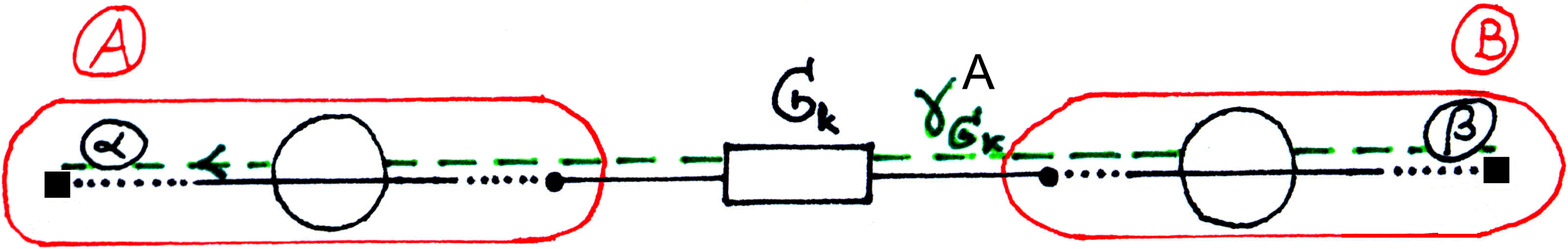}}
\vskip -1.8cm
\end{center}
The path $\gamma_{G_k}^A$ is uniquely defined as the path which starts in the local reference node ${\xy*+{{\beta}}*\cir{}\endxy}$ of ${\xy*+{\scriptstyle{B}}*\cir{}\endxy}$, passes through the node of ${\xy*+{\scriptstyle{B}}*\cir{}\endxy}$ incident with the admittance $G_k,$ along $G_k$ to that node of ${\xy*+{\scriptstyle{A}}*\cir{}\endxy}$ incident with the other end of $G_k$ and finally ends in the local reference node ${\xy*+{{\alpha}}*\cir{}\endxy}$ of ${\xy*+{\scriptstyle{A}}*\cir{}\endxy}.$ When this path is traversed in the opposite direction, we will call it $\gamma_{G_k}^{B}.$ Now let $\Sigma_k^{A}$ be the sum, relative to the orientation of this path, of those voltage sources which are part of the path $\gamma_{G_k}^{A},$ with a voltage being counted positive if the voltage source in question and the path are oriented in opposite and negative, when the orientations are the same. Consequently we have $\Sigma_k^{A}=0,$ if $G_k$ directly connects two local reference nodes.
We are now in a position to formulate the fill-in rule for the vector $\widehat{\bf J}_N:$

\smallskip
{\bf Rule 2.}
Let $\{i_m\}$ be the current sources and $\{G_k\}$ be the admittances of the circuit $\Gamma.$ Let ${\xy*+{\scriptstyle{A}}*\cir{}\endxy}$ be a supernode in $\Gamma,$ not containing the datum node. Then the entry of $\widehat{\bf J}_N$ at position $A$ 
is given by
\begin{equation}
\sum_{i_m :\, i_m \to\, {\xy*+{\scriptstyle{A}}*\cir{}\endxy}} i_m - 
\sum_{i_r :\, i_r \leftarrow\, {\xy*+{\scriptstyle{A}}*\cir{}\endxy}} i_r +
\sum_{G_k :\, G_k \,\hbox{\scriptsize incident with }{\xy*+{\scriptstyle{A}}*\cir{}\endxy}} G_k \cdot \Sigma_k^{A},
\end{equation}
where $i_m \to\, {\xy*+{\scriptstyle{A}}*\cir{}\endxy}$ means, that the current source $i_m$ is incident with ${\xy*+{\scriptstyle{A}}*\cir{}\endxy}$ and directed toward ${\xy*+{\scriptstyle{A}}*\cir{}\endxy},$ and  $i_r \leftarrow\, {\xy*+{\scriptstyle{A}}*\cir{}\endxy}$ that $i_r$ is incident with ${\xy*+{\scriptstyle{A}}*\cir{}\endxy}$ but directed away from ${\xy*+{\scriptstyle{A}}*\cir{}\endxy}.$

\smallskip
Returning to our first example, in the next figure we see the relevant paths $\gamma_{G_2}$ and $\gamma_{G_3},$ which result in the entries $G_2\cdot v_{02}$ and $G_3\cdot v_{01}$ of the right hand side of equation (\ref{exfinal}).
\begin{center}
\vskip -.05cm
{\includegraphics[width=.99\linewidth,clip,keepaspectratio]{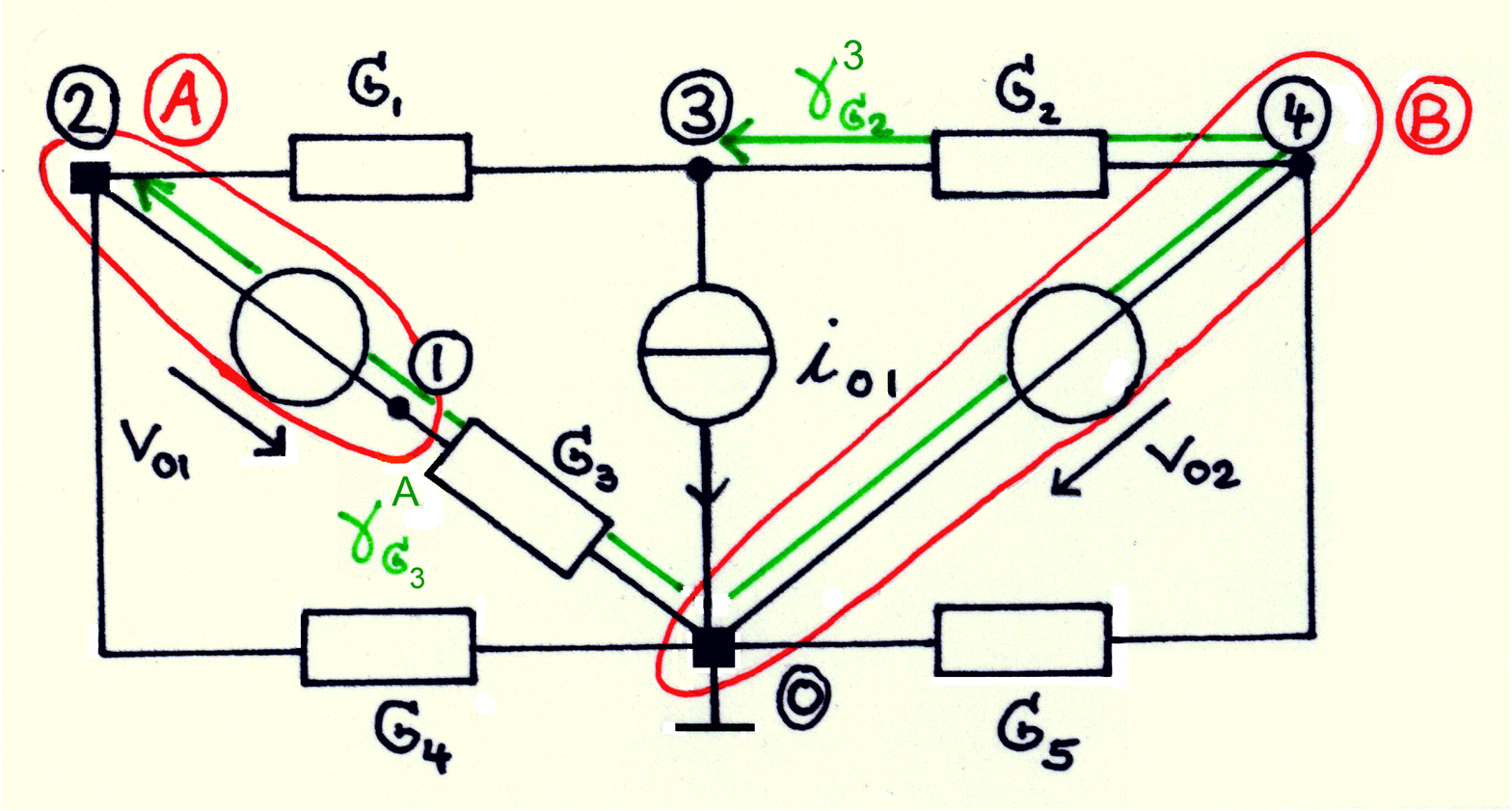}}
\end{center}
\vskip-.4truecm
For the convenience of our readers we give a second example with the supernodes marked and the relevant ''admittance paths'' sketched. 

\begin{center}
\vskip -0.20cm
{\includegraphics[width=.99\linewidth,clip,keepaspectratio]{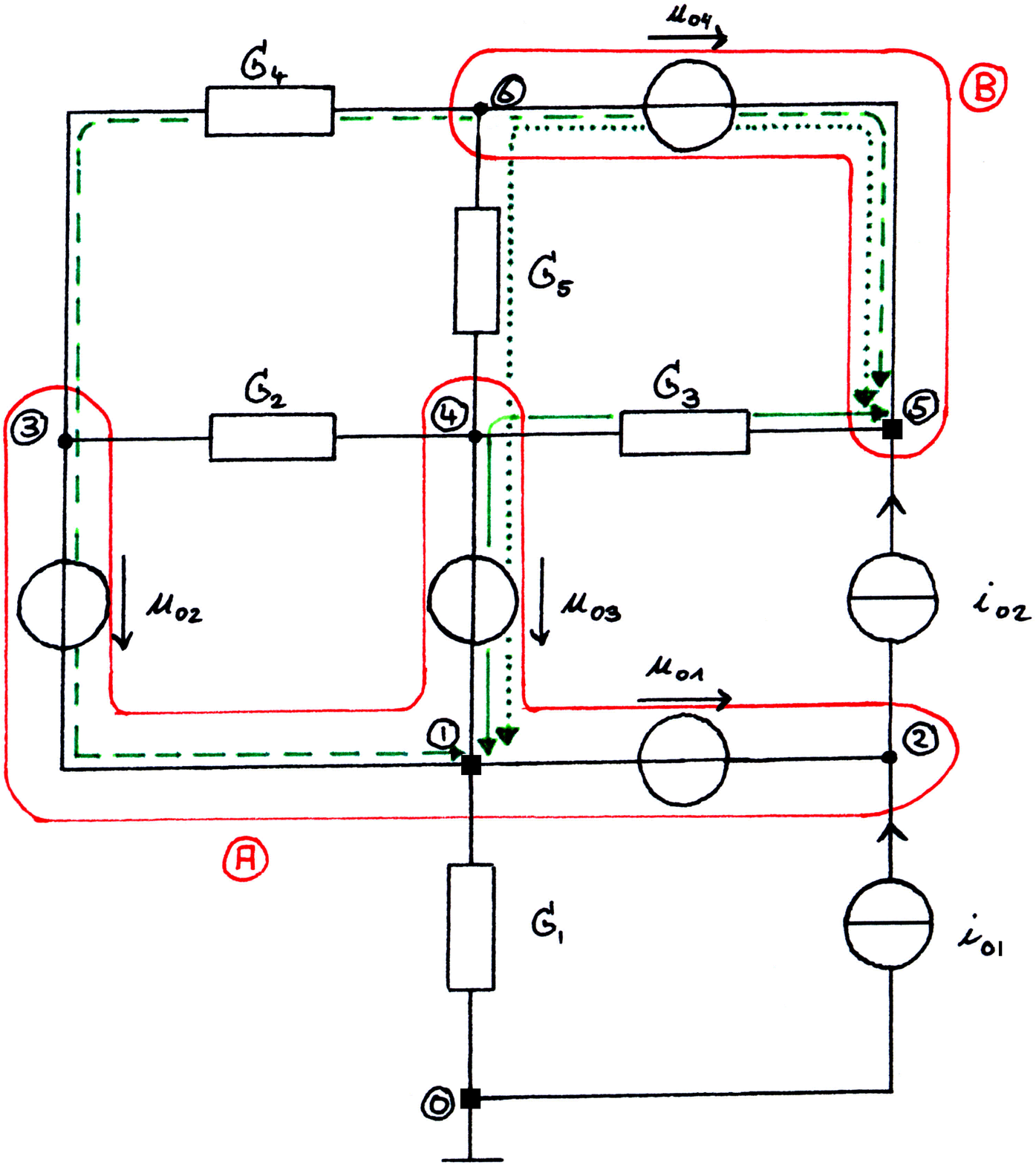}}
\end{center}
\vskip-.5truecm
Now it is easy to read off the resulting reduced system of equations. The interested reader is called upon to check each entry by using the original algorithm, to see the advantage of our approach.
\begin{equation*}
\begin{split}
&\begin{matrix}
{\xy*+{\scriptstyle{A}}*\cir{}\endxy}\\
{\xy*+{\scriptstyle{B}}*\cir{}\endxy}\\
\end{matrix}
\quad
\begin{pmatrix}
G_1+G_3+G_4+G_5 & -G_3-G_4-G_5\\
-G_3-G_4-G_5 & G_3+G_4+G_5\\
\end{pmatrix}
\cdot
\begin{pmatrix}
v_{\xy*+{\scriptstyle{1}}*\cir{}\endxy}\\
v_{\xy*+{\scriptstyle{5}}*\cir{}\endxy}\\
\end{pmatrix}
=\\
&
\begin{pmatrix}
i_{01}-i_{02} & + & G_5(u_{04}-u_{03}) + G_4(u_{04}-u_{02}) - G_3u_{03}\\
i_{02} & + & G_5(u_{03}-u_{04}) + G_4(u_{02}-u_{04}) + G_3u_{03} 
\\
\end{pmatrix}
\end{split}\end{equation*}

Although this exemplary circuit is highly artificial, it helps to stress yet another point, namely that in many circuits any try to set up nodal equations with the help of source shifting and substitution by Norton equivalent subcircuits is highly difficult or nigh impossible. 

\subsection{Circuits with Controlled Sources}
The algorithm of nodal analysis can be easily generalized to circuits containing voltage controlled current sources. Clearly supernodal analysis and the two rules given above are suitable for setting up the equations when circuits with voltage controlled sources of any kind have to be analyzed. In this case, however, the concept of a supernode has to be extended in the following way:

\begin{definition}
A subcircuit of a circuit which is connected, consists only of nodes and independent as well as the controlled branches\footnote{
The reader should keep in mind that a controlled source consists of two branches; thus we have to specify which branch should belong to the supernode.} 
of dependent voltage sources, and which is maximal with these two properties is called a {\em supernode.}
\end{definition}

By temporarily treating controlled voltage sources as independent sources (the ''taping'' of \cite{Davis}) and using this modified definition, again we can use the above algorithm plus our two fill-in rules. To complete the algorithm, in an add\-itio\-nal fourth step we need to replace the dependent voltages and currents by the controlling node voltages (i.e.\ the untaping of \cite{Davis}). In this more general case, we cannot expect a clean-cut partition of the node equations into those, describing the ''inner workings'' of each supernode and those resulting from the contracted circuit only. Never\-the\-less, our rules lead to a simple to set up interim result. 

Let us finally consider current controlled sources:
When super\-nodes have been introduced, controlling branch currents fall into two classes, those inside and those outside a supernode. Those outside are currents through admittance branches and thus can be easily substituted by the corresponding set of node voltages as variables. 
In the second case, with the controlling branch being part of a super\-node, if again we can assume that each supernode is a tree, then by induction we can show that at least one end of the branch under consideration has to be the root of a rooted tree of voltage sources. The leaves of this tree are nodes through which currents induced by admittance branches or current sources enter the supernode. Now, using KCL at each node of the rooted tree, it is easy to describe the controlling branch current as the sum of all of these currents.

\subsection{Circuits which contain Nullors}
It was the authors initial hope that the approach of supernodal analysis by inspection could somehow be made to work on circuits containing nullors as well. This hope was shattered by the example below\footnote{The reader should note the particular choice of nodes ${\xy*+{1}*\cir{}\endxy}$ and ${\xy*+{{4}}*\cir{}\endxy}$ as local reference nodes, which from a practical point of view clearly is absurd, but from an algorithmic point of view is a distinct possibility. Furthermore we have refrained from setting $v_{\xy*+{\scriptstyle{0}}*\cir{}\endxy}=0.$}.

Though for circuits containing nullors, again there is a reduced equation system which is of the form $\widehat{\bf Y}_N \widehat{\bf v}_N = \widehat{\bf J}_N$ and although there are hints of how to fill out
$\widehat{\bf Y}_N,$ in general there does not seem to be any suitable definition of the notion of a supernode which would lead to a short cut, by which the tedious setting-up of equations could be avoided: On the one hand super\-nodes in circuits with nullors should contain the norators (thus defining Kirchhoff-surfaces for which the equations have to be set up); on the other hand the voltages of nodes connected by nullators to supernodes (in the sense of definition \ref{super1}) are known from the beginning as well, and should not appear as variables in their own right. As our example shows, these two demands do not lead to any well to teach or easy to apply fill in rule.

\begin{center}
\vskip -0.2cm
{\includegraphics[width=\linewidth,clip,keepaspectratio]{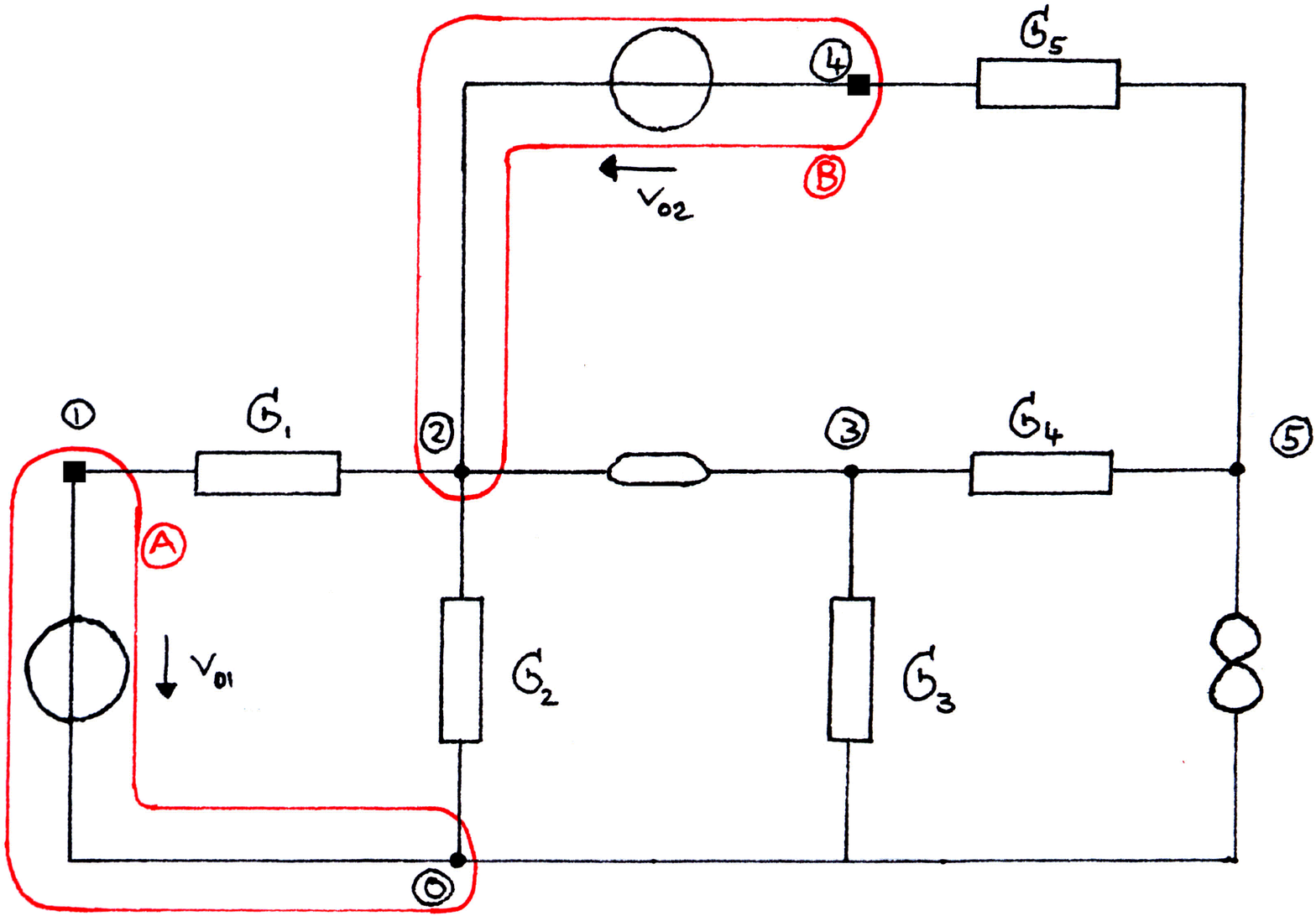}}
\vskip -0.5cm
\end{center}

\begin{equation*}
\begin{split}
\begin{matrix}
{\xy*+{\scriptstyle{B}}*\cir{}\endxy}\\
{\xy*+{{3}}*\cir{}\endxy}
\end{matrix}
\quad
\begin{pmatrix}
- (G_1+G_2) & G_1+G_2+G_5 & - G_5\\
-G_3 & G_3+G_4 & - G_4
\end{pmatrix}
\cdot
\begin{pmatrix}
v_{\xy*+{\scriptstyle{1}}*\cir{}\endxy}\\
v_{\xy*+{\scriptstyle{4}}*\cir{}\endxy}\\
v_{\xy*+{\scriptstyle{5}}*\cir{}\endxy}
\end{pmatrix}
&
=\\
\begin{pmatrix}
- G_2\cdot v_{01} + (G_1+G_2)\cdot v_{02}\\
- G_3\cdot v_{01} + (G_3+G_4)\cdot v_{02}
\end{pmatrix}
&
\end{split}\end{equation*}

 Fortunately, in case of admittance circuits with nullors, supermesh analysis or RLA, as it was presented in algorithm 3.2 and chapter 5 of \cite{SommerSuper} can be shown to be universal, and thus can be used for all circuits without cutsets of independent current sources and nullators.


\section{Conclusion}
Past experience in Braunschweig with an undergraduate course of about 150 students \cite{GerbrachtVL} has shown that the modified SNA-algorithm, with fill-in rules, as presented above, is quite easy to teach. Moreover, the students naturally adapted to it and after only a small amount of training applied it successfully to a number of problems. 
Thus it seems that this algorithm together with its counterpart, the supermesh analysis (for nullor circuits), has reached a state of maturity, that makes it perfectly suitable for presentation as ''universal tools'' in any course on circuit theory.

\bibliographystyle{IEEE}


\section*{Note added to the Electronic Version}
In this electronic document, some small typographical errors of the printed version were corrected. The figures, though still drawn by hand, now are from the original coloured drawings.

Furthermore, for the convenience of the reader the abstract has been rewritten, and keywords, ACM and MSC classifications, and a short CV according to IEEE standards have been added. One URL has been added to the bibliography.\phantom{mmmmmmmmm}\hfill 
\phantom{m}\hfill (March~9th,2009)

\begin{biography}
[{\includegraphics[width=1in,height=1.25in,clip,keepaspectratio]{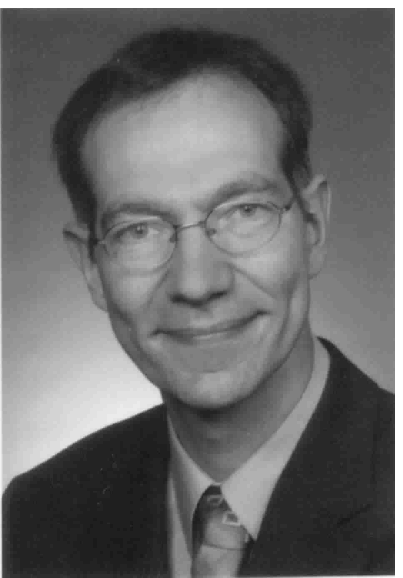}}]
{Eberhard H.-A.~Gerbracht}
received a Dipl.-Math.\ degree in mathematics, a Dipl.-Inform.\ degree in computer science, and a Ph.D. (Dr.\ rer.nat.) degree in mathematics from the Technical University Braunschweig, Germany, in 1990, 1993, and 1998, respectively.

From 1992 to 1997 he was a Research Fellow and Teaching Assistant at the Institute for Geometry at the TU Braunschweig. From 1997 to 2003 he was an Assistant Professor in the Department of Electrical Engineering and Information Technology at the TU Braunschweig. During that time he was also appointed lecturer for several courses on digital circuit design at the University of Applied Sciences Braunschweig/Wolfenb\"uttel, Germany. From 2001 to 2002 he was appointed lecturer for a two-semester course in linear circuit analysis at the TU Braunschweig. After a two-year stint as a mathematics and computer science teacher at a grammar school in Braunschweig and a vocational school in Gifhorn, Germany, he is currently working as free-lance private instructor, advisor, and independent researcher in various areas of mathematics. His research interests include combinatorial algebra, C*-algebras, the history of mathematics in the 19th and early 20th century and applications of computer algebra and dynamical geometry to graph theory, calculus, and electrical engineering.

Dr.~Gerbracht is currently a member of the German Mathematical Society (DMV), the Society of Computer Science Teachers in Lower Saxony within the Gesellschaft f\"ur Informatik (GI-NILL), and founding member of the society ``Web Portal: History in Braunschweig - \href{http://www.gibs.info}{www.gibs.info}''.
\end{biography}

\end{document}